\documentclass[12pt]{iopart}
\usepackage{iopams}
\usepackage{graphicx}
\usepackage{subfig}
\usepackage{color}         % Lines 8-14 added
\definecolor{webgreen}{rgb}{0,.5,0}

\newcommand{\Blue}{\color{blue}}

\newcommand{\bx}{\boldsymbol{x}}        
\newcommand{\by}{\boldsymbol{y}}

\newcommand{\reals}{\mathbb{R}}

\begin{document}

\title[Dynamic State Estimation from Poisson Spike Trains]{Dynamic State Estimation Based on Poisson Spike Trains - Towards a Theory of Optimal Encoding}
\author{Alex Susemihl$^{1,2}$, Ron Meir$^3$ and Manfred Opper$^{1,2}$}
\address{$^1$ Department of Artificial Intelligence, Technische Universit\"at Berlin, Franklinstr. 28-29, Berlin 10587, Germany}
\address{$^2$ Bernstein Center for Computational Neuroscience Berlin, Philippstr. 13, Haus 6, Berlin 10115, Germany}
\address{$^3$ Faculty of Electrical Engineering, Technion,  Haifa 32000, Israel}

\begin{abstract}
Neurons in the nervous system convey information to higher brain regions by the  	generation of spike trains. An important question
%which has been receiving increasing attention
in the field of computational neuroscience
% relates to the means by which
is how these sensory neurons encode environmental information in a way which may be simply analyzed by subsequent systems. Many aspects of the form and function of the nervous system have
been understood using the concepts of optimal population coding. Most studies, however, have neglected the aspect
of temporal coding. Here we address this shortcoming through a filtering theory of inhomogeneous
Poisson processes. We derive exact relations for the minimal mean squared error of the optimal Bayesian
filter and by optimizing the encoder, obtain optimal codes for populations of neurons. We also show that a class
of non-Markovian, smooth stimuli are amenable to the same treatment, and provide results for the filtering
and prediction error which hold for a general class of stochastic processes. This sets a sound mathematical
framework for a population coding theory that takes temporal aspects into account. It also formalizes a number
of studies which discussed temporal aspects of coding using time-window paradigms, by stating them
in terms of correlation times and firing rates. We propose that this kind of analysis allows for a systematic
study of temporal coding and will bring further insights into the nature of the neural code.
%
%We consider the problem of optimal population coding of neuroscience from a stochastic dynamics point
%of view. Restricting the scope of the problem we are able to derive exact solution for the MMSE of a
%Bayesian ideal observer reconstructing the driving stimulus from a noisy spike train. The MMSE
%is given by the average covariance of the posterior filtering probability. By considering the dynamics
%of the filter
%we therefore draw a number of conclusions on the MMSE in this setup. It is also straightforward to
%derive the behaviour of the prediction error. We discuss implications of this study on the study of
%optimal population codes and future extensions of this work.
\end{abstract}

\submitto{Journal of Statistical Mechanics: Theory and Experiment}

\section{Introduction}
Populations of neurons transmit information through their joint spiking activity. One of the main goals of computational neuroscience is to gain insight into the mechanisms which shape the functional activity of neurons, and to better understand and possibly decode the information encoded by neurons. In the study of optimal population codes, it is usual to start from considerations about the nature
of the neural information processing and of the tasks performed by the neurons to search for the best possible coding strategies (see for example
\cite{Bobrowski2009,Yaeli2010,Berens2011,Tkacik2010}).
Considering the encoding and decoding of stimuli from the information-theoretical viewpoint, we can develop a theory for the optimal Bayesian estimator for any given task. Given a particular estimation task and a noise model, the Bayesian estimator is the best estimator for the task, minimizing a well-defined cost function (see \cite{Robert2007}). In the broader context of Systems Theory, spike train encoding and decoding can be viewed within the context of optimal filtering based on point process observations \cite{Snyder1972}.
\par
{\Blue Here we focus on a specific question: Given a stimulus ensemble and the optimal Bayesian decoder, how can we design an {\bf encoder} to minimize the mean squared error in a stimulus reconstruction/state estimation task? That is, we would like to determine an optimal response distribution, to make the reconstruction of the stimulus easier.
%In a sense we would like to fine tune the likelihood distribution to make decoding easier.
This has been the subject of previous investigations in a number of contexts. In \cite{Brunel1998,Yaeli2010,Berens2011} the authors sought to answer the question in the framework of static stimuli. These papers established the existence of a finite optimal tuning width for bell-shaped tuning functions. In \cite{Bobrowski2009} the authors study a similar problem in the context of finite-state Markov models. In the Deep Belief Network literature, the study of Autoencoders, which deal with a very similar question, has recently received a lot of attention as well \cite{Vincent2008}. We focus here on the framework proposed in \cite{Huys2007}, where a dynamic stimulus is observed through noisy spike trains and is decoded online. This falls within the general theory of Bayesian filtering of point processes \cite{Snyder1972}. So, given an ensemble of stochastic stimuli and a family of possible encoders, we seek the encoder that minimizes the mean squared error of the Bayesian decoder (filter) in a state estimation task.}
\par

We observe that the task of designing an optimal encoder-decoder pair in our setting is very different from the case often studied within Information Theory \cite{Shannon1948}, where real-time constraints are absent and solutions are asymptotic in nature, being based on infinitely increasing block sizes. In the present real-time setting, we are interested in guaranteeing good real-time performance for {\it finite} observation times. In fact, the problem studied in this work falls into the category of a decentralized multi-agent sequential stochastic optimization problem (viewing the encoder and decoder as agents), the general solution to which is not known \cite{Mahajan2009}. The approach taken here is based on selecting a decoder using optimal Bayesian decoding assuming a fixed encoder, and then optimizing the encoder itself.\par

A central aspect of neural coding is its speed. Neural populations typically perform computations within less than $50$ milliseconds, accounting for the fast responses characteristic of animals \cite{Stanford2010,Thorpe1996}. Most studies
of optimal population coding, however, still resort to the paradigm of time-slots, in which a (mostly static) stimulus is presented to a network for a given time window, spikes are pooled for each neuron and a rate code is assumed \cite{Berens2011,Tkacik2010}. Here we follow a different path, focusing on the dynamical aspect of natural stimuli, and developing the optimal Bayesian filter for a dynamic state estimation problem (this has been hinted at in \cite{Huys2007} and developed for finite-state models in \cite{Bobrowski2009}). In filtering theory, one tries to estimate the value of a certain function of the system's state, given noisy observations of that system's state in the past.\par

Drawing from the theory of stochastic dynamics, we present a model for the joint dynamics of signal
and noise, where the signal is assumed to be a diffusion process and the noise arises from the Poisson
spiking of the neurons. For the subclass of linear stochastic processes, we find that the mean squared
error (MSE) is equal to the average posterior variance of the filter. This can be shown to be independent of the specific signal, and we can analyze the marginal dynamics of the variance of the filter. Analyzing this marginal dynamics, we
obtain results regarding the value of the mean squared error through simulations and in a mean-field approximation, which hold in the equilibrium as well as in the relaxation period. In spite of the simplifications involved, the mean-field results are shown to be in very good agreement with the the full Monte Carlo simulations of the system. \par

For the linear stochastic processes considered we obtain an interesting relation for the timescales
involved. Specifically, we find that whenever the average interspike interval is larger than the correlation
time of the observed process, the distribution of possible errors diverges around the upper bound for the error. This implies that the firing rate of the sensory neurons must exceed a threshold in order to be able to track the input properly. The threshold is defined by the statistics of the stimulus. This relation holds exactly for the class of processes considered, and is specially pronounced in the Ornstein-Uhlenbeck process, where we present a closed-form solution of the full distribution of the posterior variance. We also provide an exact analysis of the prediction error which holds for the general class of processes discussed. Furthermore, we show that for the stimuli considered there
is a finite optimal tuning width for the tuning functions which minimizes the error in the reconstruction of the stimulus. The dependence of this optimal tuning width as a function of the stimulus properties is discussed in the context of an ecological theory of sensory processing \cite{Atick1992}.
\par

{\Blue
Much effort has been devoted to understand how spiking neural networks can implement operations such as marginalization, optimal inference under uncertainty and others. The finding that humans and animals combine uncertain cues from separate senses near-optimally \cite{Ernst2002,Knill2004,Lochmann2011} has given a lot of traction to this line of research. We note that this paper takes a different path. We seek the encoder that saturates the limits given by the Bayesian optimal decoder. In \cite{Boerlin2011}, the authors have considered a similar problem, but tried to devise a spiking neural network that would optimally decode the stimulus. Similarly, in \cite{Beck2011} a procedure based on divisive normalization was presented that performs marginalization optimally for similar problems, and was applied to a filtering problem similar to the one considered here. We, however, focus on the optimal design of an encoder given an ensemble of stimuli assuming optimal decoding and study their relation to the statistical structure of the stimulus. This has been studied before in different settings (e.g. \cite{Bobrowski2009,Yaeli2010,Berens2011,Bethge2002}).
}\par

We propose that the framework of Bayesian filtering is more suited to study population coding than the usual time-binning method, as it allows for a natural inclusion of the time scales of the observed process and of the internal dynamics
of the neurons. We have shown that for a simple model, we find a general relation connecting the time
scales of the population spiking and that of the stimulus observed with the error of an ideal observer.
This has implications for the optimal tuning width of tuning functions of neurons. We believe that different strategies for temporal coding in different sensory systems might be traced to similar arguments, as has been done in \cite{Atick1992} for the limit of slow temporal stimulation.\par

The main contribution of the present paper is in providing insight, combined with closed form mathematical expressions, for the reconstruction error of stimuli encoded by biologically motivated point processes. While precise expressions can be obtained only in limiting cases, they provide an essential starting point for a theory of optimal encoding-decoding within a biologically relevant framework, and demonstrate the importance of adaptive encoding to the neural processing of dynamic stimuli. Note that even in the simple case of linear processes observed by linear systems (e.g., \cite{Anderson1979}), it is in general impossible to obtain closed form expressions for the estimation error, and one usually resorts to bounds.

\subsection{Structure of the Paper}

In section \ref{general} we present the framework used. We will derive a number of results on the Minimum Mean Squared Error (MMSE) for a stimulus reconstruction task with a Gaussian process observed through Poisson spike trains. A thorough analysis is presented for both Ornstein-Uhlenbeck processes and for smoother, non-Markovian, processes. In section \ref{optimal} we will discuss the application of these results to the study of optimal population codes. Namely we show the scaling of the optimal tuning width and its corresponding MMSE as a function of the correlation structure of the process and the overall firing rate. We finalize by discussing the implications of the presented framework to the field and future applications of our framework.

\section{Reconstructing Dynamic Stimuli Based on Point Process Encoding}
\label{general}

The problem of reconstructing dynamic stimuli based on noisy partial observations falls within the general field of {\it filtering theory} (e.g., \cite{Anderson1979}). Consider a dynamic stimulus $\bx(t)$, $\bx \in \reals^n$, which is observed through a noisy set of sensors leading to output $\by(t)$, $\by\in\reals^m$. For example, $\bx(t)$ could represent the position and velocity of a point object, and $\by(t)$ could represent the firing patterns of a set of retinal cells. The objective is to construct a filter, based only on the observation process, which provides a good estimator for the unknown value of the stimulus. Formally, denoting by $\by([0,t])$ the set of observations from time $0$ to the present time, we wish to construct an estimator $\boldsymbol{\hat x}(\by([0,t]))$ which is as close as possible (in some precisely defined manner) to $\bx(t)$. A classic example of filtering is the case of a stimulus $\bx(t)$ generated by a noisy linear dynamical system, and an observer $\by(t)$ which is based on a noisy linear projection of the stimulus. The classic Kalman filter (e.g., \cite{Anderson1979}) then leads to the optimal reconstruction in the MMSE sense.\par

{\Blue For a filtering task, the use of an $\mathcal{L}_p$ norm as a cost function is a natural choice, given that we are interested in reconstructing the system's state as precisely as possible. Here we choose to use the $\mathcal{L}_2$ norm as a cost function for our reconstruction task. This is a natural choice for a filtering task (see \cite{Anderson1979}) and has been frequently used in studies of optimal population coding \cite{Bobrowski2009,Yaeli2010,Berens2011}. 
Another popular cost function for studies of optimal population coding is the mutual information between the input distribution and the conditional response distribution \cite{Tkacik2010}. This allows one to find the code that optimally codes the information contained in the input in its response. Though recent theoretical advances are sketching out the relationship between information- and MMSE-optimal codes \cite{Palomar2007,Merhav2011,Guo2008}, there seems to be no simple equivalence between these two cost functions. In Gaussian additive channels, the MMSE is equal to the derivative of the mutual information between input and output with respect to the signal-to-noise ratio. Though similar relationships have been derived for Poisson processes, these hold only for linearly modulated inhomogeneous processes and do not relate directly to the MMSE \cite{Guo2008,Atar2011}. Furthermore, these results have been derived only for single point processes, not for populations thereof. We therefore choose to work strictly with the $\mathcal{L}_2$ cost function, as this is not only a natural choice for the problem at hand, but also allows for a number of analytical results.}
\par
We consider the case of linear Gaussian stochastic processes observed by a population of neurons with unimodal tuning functions. This is analogous to considering stimuli drawn from a Gaussian process prior, as has been done in \cite{Huys2007}. We will have a prior distribution over the stimulus $x(t)$ given by a Gaussian process with zero mean and covariance function $K(t,t')=\left<x(t) x(t')\right>$,  where the angled brackets denote the average over the ensemble of Gaussian processes.\par

Rather than considering general Gaussian processes we will focus on a class of processes which are particularly amenable to analytic investigation. This will allow us to consider both simple Markov Gaussian processes (the so-called Ornstein-Uhlenbeck process) and higher order Markov Gaussian processes.
We will consider stochastic processes described by a stochastic differential equation of the form
\begin{equation}
\left(\frac{\rmd}{\rmd t} + \gamma\right)^P x(t) = \eta \frac{\rmd W(t)}{\rmd t},
\label{eqn:langevin_eqn}
\end{equation}
where $W(t)$ is a scalar Wiener process. We can find the covariance of the process by
calculating the Fourier transform, computing the power spectrum and then reversing the Fourier transform.
We will have
\begin{equation}
\left<\tilde{x}(\omega) \tilde{x}^*(\omega)\right> = \frac{\eta^2}{(\gamma+2 \pi \rmi\omega)^P(\gamma-2 \pi \rmi\omega)^P}.
\end{equation}
This power spectrum leads to stochastic processes with the so-called Matern kernel \cite[p. 211]{Rasmussen2005}. If unobserved, the distribution over $x$ will converge to the equilibrium distribution, given by a Gaussian distribution with zero mean and covariance $\sigma^2_x = \eta^2/2^P \gamma^{2 P -1}$.
These
processes can also be written as multidimensional first-order processes by defining $\mathbf{X}(t) \equiv \left(\mathbf{X}_1(t) ,\mathbf{X}_2(t) ,\ldots, \mathbf{X}_P(t)\right)^\top= \left(x(t),x^{(1)}(t),\ldots, x^{(P-1)}(t)\right)^\top$, where $x^{(i)}(t)$ denotes the $i$-th derivative of of $x(t)$, and its associated stochastic differential equation
\begin{equation}
\label{eqn:vector_sde}
\rmd\mathbf{X}(t) = -\Gamma \mathbf{X}(t) \rmd t+H \rmd\mathbf{W}(t),
\end{equation}
where $\mathbf{W}(t)$ is a $P$-dimensional Wiener process. $\Gamma$ and $H$ are defined in \ref{sec:app_matrix}. Note that the process $x(t)$ itself will not in general be Markovian. The process $\mathbf{X}(t)$ of $x(t)$ along with
its first
$P-1$ derivatives will, however, be Markov. This allows us to treat smooth non-Markov dynamics as the
marginal case of a higher-dimensional Markov stimulus and so to draw from
the theory of Markov stochastic processes, which is very well-established \cite{Gardiner2004}.
\par

The spike trains will be modeled by inhomogeneous Poisson processes $N^m(t)$, namely, Poisson processes
whose rates are functions of the stochastic process $\mathbf{X}(t)$. More precisely, $N^m(t)$ represents
the number of times neuron $m$ has spiked since the beginning of the experiment.
Furthermore, we will assume each spike train $N^i(t)$ to be conditionally independent of all others, that is, given a value of the stimulus $\mathbf{X}(t)$ the
spiking of each neuron is independent of all others. The function relating the stimulus $\mathbf{X}(t)$ to
the rate of the Poisson process is often referred to
as a tuning function, and will be denoted by $\lambda_m(\mathbf{X}(t))$. We will consider unimodal tuning function of the form
$$
\lambda_m(x(t)) =\phi \exp\left(-\frac{(x(t) - \theta_m)^2}{2\alpha^2}\right),
$$
where $\theta_m$ is the preferred stimulus value of neuron $m$. Tuning functions of this form are often found in orientation-selective cells in visual cortex of mammals and in place cells in the hippocampus \cite{Benucci2009,Okeefe1971,Knierim1995}. For multi-dimensional stimuli, we can write these more generally as
$$
\lambda_m(\mathbf{X}(t)) =\phi \exp\left( -(\mathbf{X}(t)-\Theta_m)^\top A^+ (\mathbf{X}(t)-\Theta_m)/2\right).
$$
{\Blue We will prefer this notation as it allows us to derive a general theory which also holds for multidimensional stimuli. The case of a one-dimensional $P$-th order process along with its $P-1$ derivatives would be recovered by setting} $A^+_{i,j} = \delta_{1,i}\delta_{1,j}/\alpha^2$ and $\Theta_m = (\theta_m,0,\ldots,0)$. While in many cases biological tuning functions are unimodal, we have chosen to work with Gaussian functions for reasons on analytic tractability.\par

The likelihood of a spike train $\{N^m([0,t])\}$ is given by (see \cite{Snyder1991})
$$
\mathcal{L}(\{N^m([0,t])\}|\mathbf{X}([0,t])) = \exp\left(\sum_m \int \rmd N^m(t) \log(\lambda_m(\mathbf{X}(t)))-\sum_m\int \rmd t\,  \lambda_m(\mathbf{X}(t))\right).
$$
We have denoted here by $\{N^m([0,t])\}$ the value of all spiking processes $N^m(s)$ for any instant $s$ such that $0 \leq s \leq t$, and likewise $\mathbf{X}([0,t])$ denotes all values of $\mathbf{X}(s)$, for all $s$ such that $0\leq s\leq t$.
With this likelihood we can then find the posterior distribution for $\mathbf{X}([0,t])$ using Bayes' rule,  
$$
P(\mathbf{X}([0,t])|\{N^m([0,t])\}) \propto\mathcal{L}(\{N^m([0,t])\}|\mathbf{X}([0,t])) P(\mathbf{X}([0,t])).
$$
Averaging out the values of $\mathbf{X}(t)$ up to but excluding the time $t$, will give us the
time-dependent Gaussian posterior $P(\mathbf{X}(t)|\{N^m([0,t])\})$. If the likelihood were Gaussian
in $\mathbf{X}$ we could use results from conditional distributions on Gaussian measures to average
over $\mathbf{X}([0,t))$ (for more
details, see \cite{Huys2007}). This is indeed the case when the tuning functions are densely packed, as we will see in the following.\par
Let us then turn to the spiking dynamics of the whole population. Because each neuron spikes as a Poisson
process with rate $\lambda_m(\mathbf{X}(t))$, and the processes $N^m(t)$ are conditionally
independent when
conditioned on $\mathbf{X}(t)$, the process $N(t) = \sum_m N^m(t)$ is also Poisson with rate
$\lambda(\mathbf{X}(t)) \equiv \sum_m \lambda_m(\mathbf{X}(t))$.
If the neurons' tuning functions are dense, however, the sum does not depend on $\mathbf{X}(t)$ and
therefore the overall firing rate of the population does not depend on the stimulus at all. The rate can
be estimated for equally spaced tuning centers $\Theta_m$ by considering it a Riemann sum \cite{Yaeli2010}. Assuming the tuning centers $\Theta_m$ are tiled in a regular lattice with spacing
$\Delta \Theta$ along each axis, we have
%$$
%1 = \int_{S(A)} \, \rmd\Theta \, \mathcal{N} (\Theta; \mathbf{X}(t), A)\approx \frac{|\Delta\Theta|^n}{\left[(2\pi)^{rank(A)} \det^*(A)\right]^{1/2}} \sum_m \exp\left(-(\Theta_m - \mathbf{X}(t) A^{+} (\Theta_m-\mathbf{X}(t))/2\right),
%$$
%and therefore
$$
\lambda = \phi \sum_m\exp\left(-(\Theta_m - X)^\top A^{+} (\Theta_m-X)/2\right) \approx \frac{\phi
\left[(2\pi)^{rank(A)} \det^*(A)\right]^{1/2}}{|\Delta\Theta|^n},
$$
where $\det^*(A)$ is the
pseudo-determinant of $A$ defined in \ref{sec:app_determ}.
The assumption of dense tuning functions is clearly very strong as the number of neurons necessary to cover an $n$-dimensional stimulus space grows exponentially with $n$.
Note however, that the deviation from this approximation is very small when the tuning center spacing is
of the order of the diagonal elements of $A$, i.e. when the tuning functions have a strong overlap. If this assumption is violated we would have to treat the
filtering problem through the stochastic partial differential equation for the probability, as was done in \cite{Snyder1972} for general Poisson processes.
\par
Given the assumption of dense tuning functions and due to the prior assumption about $x(t)$, the likelihood becomes Gaussian, and the posterior distribution $P(\mathbf{X}(t)|\{N^m([0,t])\})$ will also be Gaussian. Since the dynamics of $\mathbf{X}(t)$ is linear, the Gaussian distribution will be conserved and, in the absence of spikes, the mean $\mu(t)$ and covariance $\Sigma(t) = \left<(\mathbf{X}(t)-\mu(t)) (\mathbf{X}(t)-\mu(t))^\top\right>$ will evolve as
\numparts
\begin{equation}
\label{eqn:free_mean_dynamics}
\frac{\rmd\mu(t)}{\rmd t}=-\Gamma \mu(t),
\end{equation}
and
\begin{equation}
\label{eqn:free_var_dynamics}
\frac{\rmd\Sigma(t)}{\rmd t} = -\Gamma\Sigma(t) -\Sigma(t) \Gamma^\top + H^2.
\end{equation}
\endnumparts
If a spike from a neuron $m$ occurs at a time $t$, the posterior distribution of $\mathbf{X}(t)$ gets updated via Bayes' rule. The prior is then given by $\mathcal{N}(\mathbf{X}(t);\mu(t),\Sigma(t))$ and the likelihood is $\mathcal{N}(\mathbf{X}(t);\Theta_m,A)$. We denote the mean and covariance immediately after the spike at time $t$ by $\mu(t^+)$ and $\Sigma(t^+)$.
By standard Gaussian properties we obtain 
\numparts
\begin{eqnarray}
\label{eqn:mean_matrix_jump}
\mu(t^+) &= \left(\Sigma(t)^{-1} + A^+\right)^{-1}(\Sigma(t)^{-1}\mu(t)+A^+ \Theta_m)\nonumber\\
&= \mu(t) +  \left(\Sigma(t)^{-1} + A^+\right)^{-1}A^+ \left(\Theta_m - \mu(t)\right)
\end{eqnarray}
\begin{eqnarray}
\label{eqn:var_matrix_jump}
\Sigma(t^+) &=\left(\Sigma(t)^{-1}+A^+\right)^{-1}\nonumber\\
&= \Sigma(t) + \left(\Sigma(t)^{-1}+A^+\right)^{-1}A^+\Sigma(t)
\end{eqnarray}
\endnumparts
This fully determines the optimal Bayesian filter, namely, it is given by $P_t(\mathbf{X}) = \mathcal{N}(\mathbf{X};\mu(t),\Sigma(t))$, where the evolution of $\mu(t)$ and $\Sigma(t)$ is given by \eref{eqn:free_mean_dynamics} and \eref{eqn:free_var_dynamics} in the absence of spikes and by
\eref{eqn:mean_matrix_jump} and \eref{eqn:var_matrix_jump} whenever there are spikes. It is interesting to observe that the dynamics of the mean and variance between spikes, given in \eref{eqn:free_mean_dynamics} and \eref{eqn:free_var_dynamics}, is precisely that obtained for the continuous time Kalman filter \cite{Anderson1979} when observations are absent. This is not surprising  since between spikes we are tracking a linear dynamical system, as does the Kalman filter.\par

%{\Blue To define an optimal estimator, one needs to define a loss function for the estimation task at hand. We will choose to work with the classical squared error loss function for the remainder of the paper. First, let us note that, since we focus on a stimulus reconstruction task, a $L$
%}\par

In order to determine the properties of the optimal encoder, we aim at obtaining expressions for the mean-squared error of the optimal filter. This gives us a measure of the best-case performance that can be achieved for a signal reconstruction task from the observation of the spike train. Specifically, we want a measure of the average performance over all stimuli and all spike trains.
For that, note that because of the Gaussian nature of the stimulus and the linear dynamics of the prior, the evolution
of the posterior variance does not depend on the spike trains of each neuron, but rather only
on $N(t)$, the total spike count of the population. The average of the posterior covariance matrix is given by
$$
\mathcal{S} (t) = \left<(\mathbf{X}(t)-\mu(t))(\mathbf{X}(t)-\mu(t))^\top\right>_{\mathbf{X},\{N^m([0,t])\}}.
$$
{\Blue
Note that the diagonal terms give us the mean squared error on the estimation of every coordinate of $\mathbf{X}(t)$. More specifically, the MMSE in the estimation of $x(t)$ is given by $\mathcal{S}_{11}(t)$.}
We can simplify this by noting that $\mu(t) = \mathbf{E}(\mathbf{X}(t)|\{N^m([0,t])\})$ and evaluating the average over
$P(\mathbf{X}(t)|\{N^m([0,t])\})$ \cite{Yaeli2010}. We have
$$
\mathcal{S} (t) = \left<\Sigma(t)\right>_{\{N^m([0,t])\}} = \left<\Sigma(t)\right>_{N([0,t])},
$$
where in the last step we have used the fact that the dynamics of $\Sigma(t)$, and therefore its average, only depend on the population spike count
 $N(t)$. Up to the last step, this derivation is generally valid, though one must take care to consider the marginal distribution of the observations, averaging out the signal.
Note that, although we can give an account of the temporal evolution of the mean squared error this cannot be given a biological interpretation in the absence of an experiment to contextualize the time dependence of the error. {\Blue However, in the limit of long spike trains, the MSE will converge to its  equilibrium value (see \fref{fig:1} and \fref{fig:3}) and we can study the equilibrium statistics of $\Sigma(t)$ to determine the average MSE of the reconstruction task.}\par
We will look at the transition probability for $\Sigma(t)$ after marginalization over $N(t)$.
%with respect to the population spiking activity $N(t)$.
Consider the probability $P(\Sigma', t+dt|\Sigma, t)$, of the covariance having a value
$\Sigma'$ at time $t+dt$ given that at time $t$ the covariance was $\Sigma$. For
infinitesimal $dt$, $\Sigma'$ is either given by  \eref{eqn:free_var_dynamics}, or with a
probability $\lambda\, dt$ there is a jump as specified in  \eref{eqn:var_matrix_jump}, with
$\lambda = \sum_m \lambda_m$. This will yield the transition probability
\begin{eqnarray}
P(\Sigma', t+dt|\Sigma, t) &=  (1-\lambda\,dt) \delta\left( \Sigma'-\Sigma -dt( H^2-\Gamma\Sigma-\Sigma\Gamma^T)\right)\nonumber \\&+ \lambda\,dt\, \delta\left(\Sigma'-(\Sigma^{-1}+A^+)^{-1} \right).
\label{eqn:transition_prob_matrix}
\end{eqnarray}
The evolution of the time-dependent probability over the covariance matrices  $P(\Sigma,t)$ will be given by the Chapman-Kolmogorov equation \cite[p.47]{Gardiner2004}
\begin{equation}
\frac{\partial P(\Sigma,t)}{\partial t} = -\nabla \left[B(\Sigma) P(\Sigma,t)\right] + \lambda C(\Sigma) P\left( (\Sigma^{-1} - A^+)^{-1},t\right) - \lambda P(\Sigma,t),
\label{eqn:DCKE_multidimensional}
\end{equation}
where $B(\Sigma) = -\Gamma\Sigma -\Sigma \Gamma^\top + H^2$. Note that this gives us the $MSE(t)$ through the average of $\Sigma$ over the distribution $P(\Sigma,t)$.
The term $C(\Sigma)$ arises from the integration of the second Dirac delta function in \eref{eqn:transition_prob_matrix} and is given by $C(\Sigma) =
1/|\det(J(\Sigma))|$, where $J(\Sigma)$ is the Jacobian matrix
$$
J_{(i,j),(k,l)} = \frac{\partial (\Sigma^{-1}+A^+)^{-1}_{i,j}}{\partial \Sigma_{k,l}} = (I+A^+\Sigma)^{-1}_{k,i}(I+\Sigma A^+)^{-1}_{j,n}
$$
The exact choice of the ordering of the indices is not of importance, as it would only account for a change in the sign of the determinant in $C(\Sigma)$, which only enters the equation through its absolute value. This
term is not of great importance, however, as it will be cancelled when we calculate the evolution of the
averages.\par
We can now formalize what we mean by the equilibrium condition mentioned above. Note that under the stimulus and noise models proposed the evolution of the distribution of the error is given by \eref{eqn:DCKE_multidimensional}. Given some initial condition for $P(\Sigma,t_0)$, the distribution will evolve and eventually it will reach an equilibrium, such that $\partial_t P(\Sigma,t) = 0$. We are interested in this equilibrium regime, as it provides the average performance of the optimal filter after all transients from the initial conditions have vanished.\par
The evolution of $\mathcal{S}(t) = \left<\Sigma\right>_t$ can be found easily from \eref{eqn:DCKE_multidimensional}. Using
$$
\frac{\rmd}{\rmd t} \int \rmd \Sigma f(\Sigma) P(\Sigma,t)= \int \rmd \Sigma f(\Sigma) \frac{\partial P(\Sigma,t)}{\partial t},
$$
and integrating by parts we obtain
\begin{equation}
\label{eqn:average_matrix_dyn}
\frac{\rmd\left<\Sigma\right>_t}{\rmd t} = -\Gamma \left<\Sigma\right>_t -\left<\Sigma\right>_t \Gamma^\top + H^2 +\lambda\left<\left(\Sigma^{-1}+A^+\right)^{-1}A^+\Sigma\right>_t,
\end{equation}
where $\left<f(\Sigma)\right>_t = \int \,\rmd\Sigma\, f(\Sigma) P(\Sigma,t)$.
This cannot be treated exactly, though, as the nonlinear averages on the right hand side are intractable even for simple cases. Similar
relations for the evolution of the average mean and variance of a filter were presented in \cite{Snyder1972}
as approximations to the true filter (based on discarding higher order moments).
\par

The evolution of the distribution over covariance matrices is thus determined. However, evaluating the averages is
intractable even for the case of dense neurons when $\lambda$ is independent of $\mathbf{X}(t)$. We will therefore look into three different ways of treating \eref{eqn:DCKE_multidimensional}. First, we can simply simulate the
population spiking process $N(t)$ as a Poisson process to obtain samples of $P(\Sigma,t)$. By
averaging over multiple realizations of $N(t)$ we can estimate the average value of $\Sigma(t)$ and by
monitoring its evolution we can determine when it has relaxed to the equilibrium. The second possibility
is to analyze  \eref{eqn:average_matrix_dyn} in a mean-field approximation, i.e. to simply replace all averages of the form $\left<f(\Sigma)\right>$ with $f(\left<\Sigma\right>)$, disregarding all fluctuations around the mean.
 We would then have the dynamics
\begin{equation}
\label{eqn:average_matrix_eq_mf}
 \frac{\rmd \left<\Sigma\right>_t}{\rmd t}= -\Gamma \left<\Sigma\right>_t -\left<\Sigma\right>_t \Gamma^\top + H^2 +\lambda\left(\left<\Sigma(t)\right>_t
 ^{-1}+A^+\right)^{-1}A^+\left<\Sigma(t)\right>_t.
\end{equation}
A third possibility is to analyze \eref{eqn:DCKE_multidimensional} directly. This leads to a number of interesting results for the one-dimensional Ornstein-Uhlenbeck process, but the generalization to higher dimensions is not straightforward. We
will proceed by analyzing the case of the Ornstein-Uhlenbeck process first. Subsequently we will
look at the case of smoother linear stochastic processes, which are produced by Gaussian processes
with Matern kernels \cite{Rasmussen2005}.
The generalization to linear diffusion processes is straightforward. We will finalize this section with a discussion of the prediction error for the optimal filter considered.\par

\subsection{Filtering the Ornstein-Uhlenbeck Process}

\begin{figure}
\begin{center}
\includegraphics[width=\columnwidth]{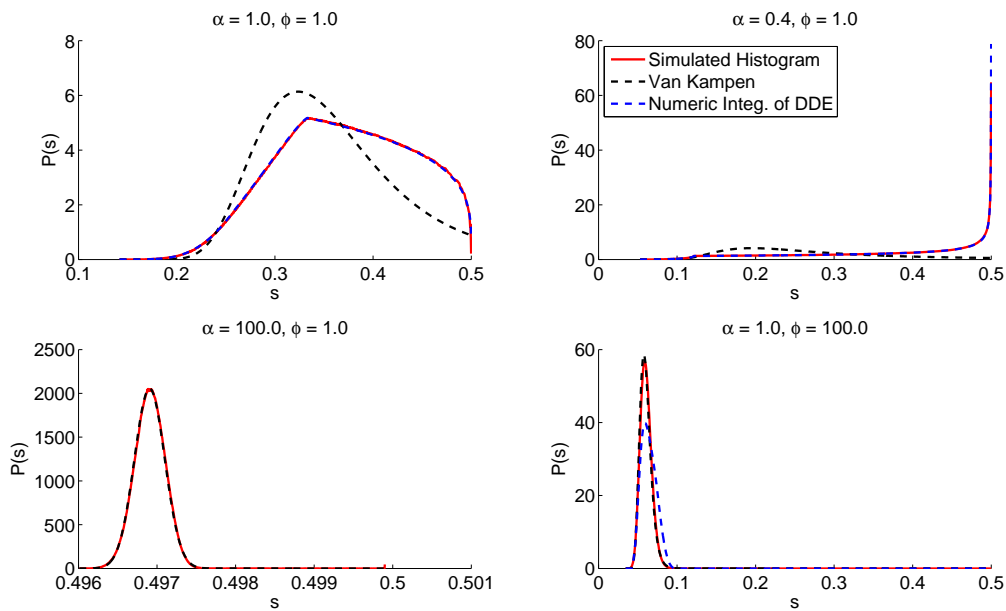}
\caption{Different treatments of the distribution of the variance for different sets of parameters of the encoder for a first-order OU stimulus. The red line shows the histogram obtained through simulations. The blue line shows the numerical integration of \eref{eqn:DCKE_1d_eq} using the mentioned boundary condition. The black line shows the Van Kampen expansion given in the text. Note that the numerical integration of \eref{eqn:DCKE_1d_eq} provides a very good account of the simulated histogram, except for very large values of $\alpha$, when the integration becomes unstable. The Van Kampen approximation in turn provides a good account for the limit of large firing rates, for large $\alpha$ as well as for large $\phi$.}
\label{fig:2}
\end{center}
\end{figure}

As is well known, the OU process is the only homogeneous Markov Gaussian process in one dimension, and is therefore particularly convenient as a starting point for analysis \cite{Gardiner2004}.
When we consider the OU process described by the stochastic differential equation
$$
\rmd x(t) = -\gamma x(t) \rmd t+ \eta \, \rmd W(t),
$$
the analysis we presented above is greatly simplified. The evolution of the posterior variance $ s$ is
then given simply by
\begin{equation}
\frac{\rmd\left< s\right>_t}{\rmd t} = -2\gamma \left< s\right>_t + \eta^2 -\lambda\left< \frac{ s^2}{\alpha^2+ s}\right>_t.
\end{equation}
We will denote the one-dimensional variance by $s$, reserving $\Sigma$ for the multidimensional covariance matrix.
In the case of the OU process we can give a more complete account of the distribution of the
errors for the filter. For the one-dimensional case \eref{eqn:DCKE_multidimensional} simplifies to
\begin{equation}
 \frac{\partial P(s,t)}{\partial t} = \frac{\partial}{\partial s}\left[(2\gamma s -\eta^2) P(s,t)\right] + \lambda \left(\frac{\alpha^2}{\alpha^2-s}\right)^2 P\left(\frac{\alpha^2 s}{\alpha^2-s},t\right) - \lambda P(s,t).
\label{eqn:DCKE_onedimensional}
\end{equation}
Clearly $P(s,t) = 0, \forall s <0$.
In the equilibrium we will have
\begin{equation}
-\frac{\rmd}{\rmd s}\left[(2\gamma s -\eta^2) P(s)\right] = \lambda \left(\frac{\alpha^2}{\alpha^2-s}\right)^2 P\left(\frac{\alpha^2 s}{\alpha^2-s}\right) - \lambda P(s).
\label{eqn:DCKE_1d_eq}
\end{equation}
This is a delayed-differential equation with nonlinear delays in $s$. {\Blue To see this, note that for every
$s $ such that $0<s<\alpha^2$, we have $\alpha^2s/(\alpha^2-s)>s$. For $s>\alpha^2$, this term becomes meaningless, as the posterior variance will always fall below $\alpha^2$ after the observation of a spike.}  This means, that for any $s>0$, the
derivative of $P(s)$ is a function of $P(s)$ and $P(\alpha^2s/(\alpha^2-s))$.
Defining the function
$j(s) = \alpha^2s/(\alpha^2+s)$ we can define intervals $I_0 = [j( s_0), s_0), \,
I_1 = [j^2( s_0),j( s_0)), \ldots$ with $ s_0 = \eta^2/2\gamma$. We will then have that given
the solution of $P( s)$ on $I_n$, the solution in $I_{n+1}$ is the solution of a simple inhomogeneous
ordinary differential equation with a continuity condition on $j^n( s_0)$. This can be simply solved
through numerical integration schemes.
We will show in \ref{sec:app_bound} that in the equilibrium $P( s>\eta^2/2\gamma) = 0$.
We therefore define the boundary conditions as $P_{eq}( s) =0,\, \forall  s>\eta^2/2\gamma$. This will imply that the jump term in  \eref{eqn:DCKE_1d_eq}
will be absent whenever the jump originates in the domain $ s>\eta^2/2 \gamma$. This is the
case when $ s \in I_0$. For these values
of $ s$ we arrive at
\begin{equation}
\frac{\rmd}{\rmd  s}\left[(2\gamma  s-\eta^2) P_{eq}( s)\right]  - \lambda P_{eq}( s) = 0.
\end{equation}
This is solved by
\begin{equation}
\label{eqn:eq_dist_sigma}
P_{eq} ( s) = C \left(\frac{\eta^2}{2\gamma}- s\right)^{\frac{\lambda}{2\gamma} -1},\quad s \in I_0.
\end{equation}
We can use this solution as a boundary condition to solve  \eref{eqn:DCKE_1d_eq} as a
delayed-differential equation with variable delays, but an analytical solution is not available for the
subsequent intervals. In \fref{fig:2} we present the numerical solution of
\eref{eqn:DCKE_1d_eq} alongside with histograms from simulations. Note that the solution in
\eref{eqn:eq_dist_sigma} is exact for the interval $I_0$ and is therefore in significant agreement with the
simulations. In the subsequent intervals numerical errors degrade the precision of the solution. Especially
when $\alpha$ is very large the intervals $I_n$ will get smaller, and the numerical integration
of \eref{eqn:DCKE_1d_eq} will become less stable.\par
This solution can also be
obtained in the limit of very small firing rates, that is when $\lambda \ll 2 \gamma$. We can then assume
that between two spikes the variance has already relaxed to its free equilibrium value
$ s_0$. A spike at time $t_s$ will then take the variance to
$ s' =j( s_0)$. The evolution will then be given by
$$
 s(t) = \rme^{-2\gamma (t-t_s)}  s' +  s_0 (1-\rme^{-2\gamma (t-t_s)}),
$$
which can be inverted into
$$
t - t_s= -\frac{1}{2\gamma}\log \left(\frac{ s_0- s(t)}{ s_0 -  s'}\right).
$$
We can then write the distribution over $ s$ as a distribution over interspike intervals
$\tau = t-t_s$.
We know the probability distribution over $\tau$ is given by an exponential with coefficient $\lambda$.
Changing the variables, we will have
$$
P( s) = P(\tau) \left|\frac{\rmd\tau}{\rmd s}\right|  \propto \rme^{-\lambda\tau} \rme^{2\gamma\tau}.
$$
Which in turn gives us  \eref{eqn:eq_dist_sigma} for the equilibrium distribution of $ s$. Note
that this derivation only holds in the limit of $\lambda \ll 2\gamma$, while the deduction above for the
interval $[ s', s_0)$ holds irregardless of the value of the parameters. The mentioned limit is
very useful in the multidimensional case, however.
\par

\begin{figure}
\includegraphics[width=\columnwidth]{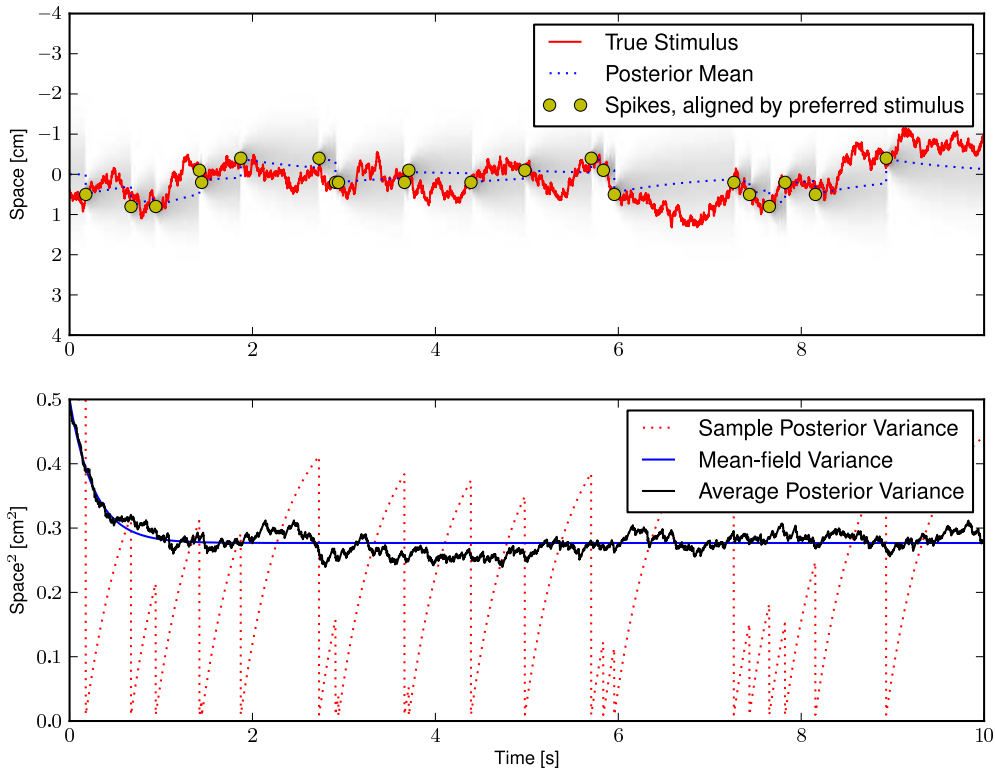}
\caption{General setup of the inference task. The upper panel shows the true stimulus or signal $x(t)$ in red solid line, the filtered posterior mean in the blue dotted line and the shading represents the posterior probability distribution. The lower panel shows the evolution of the posterior variance. The red dotted line gives the sample variance of the sample in the upper panel, the solid black line shows an average over 100 realizations and the blue line gives the evolution of the mean-field equation with the same initial conditions.}
\label{fig:1}
\end{figure}

In the mean-field approximation for the equilibrium
condition we will have
\begin{equation}
\frac{\rmd s}{\rmd t}=-2\gamma  s + \eta^2 - \lambda \frac{ s^2}{\alpha^2+ s}.
\end{equation}
This gives a remarkably good account of the equilibrium properties of the system, as is show in 
\fref{fig:1}. 
%The MSE shows a pronounced minimum for a given value of the tuning width $\alpha$ and decays to $0$ as the overall firing rate goes to infinity.
Surprisingly, the mean-field
equation gives a very good account of the transient behavior of the error as well as of the equilibrium.\par
We can also derive a simple Gaussian approximation for the rescaled inverse variance
$z = \eta^2/(\gamma s)$. Note
that the distribution of the inverse variance at the equilibrium is given by
\begin{equation}
-\frac{\rmd}{\rmd z}\left[\gamma z(2-z) P_{eq}(z)\right] + \lambda P_{eq}(z - \frac{\eta^2}{\gamma\alpha^2}) - \lambda P_{eq}(z) =0.
\label{eqn:zode}
\end{equation}
In the limit where $\eta^2/\gamma\alpha^2 \ll z$ we can Taylor expand the second term to second order. In this regime we will have very broad tuning functions and each individual spike will have little effect. This can be thought of as a diffusion limit. Further linearizing the drift term around its equilibrium we will obtain the Van Kampen expansion for the problem \cite{Gardiner2004}. It will lead to a Fokker-Planck equation whose solution is
the Gaussian distribution $P_z(z) = \mathcal{N}\left(1+\left[1+\lambda\delta/\gamma\right]^{1/2},\lambda
\delta^2\left[4\gamma\right]^{-1}\left[1+\lambda\delta/\gamma\right]^{-1/2}\right)$. By writing $P( s) =
P_z(\eta^2/\gamma s)\eta^2/\gamma s^2$ we can then recover the distribution over the variances. These results are all compared in \fref{fig:2}.

\label{sec:ou}

\subsection{Filtering Smooth Processes}

\begin{figure}
\includegraphics[width=\columnwidth]{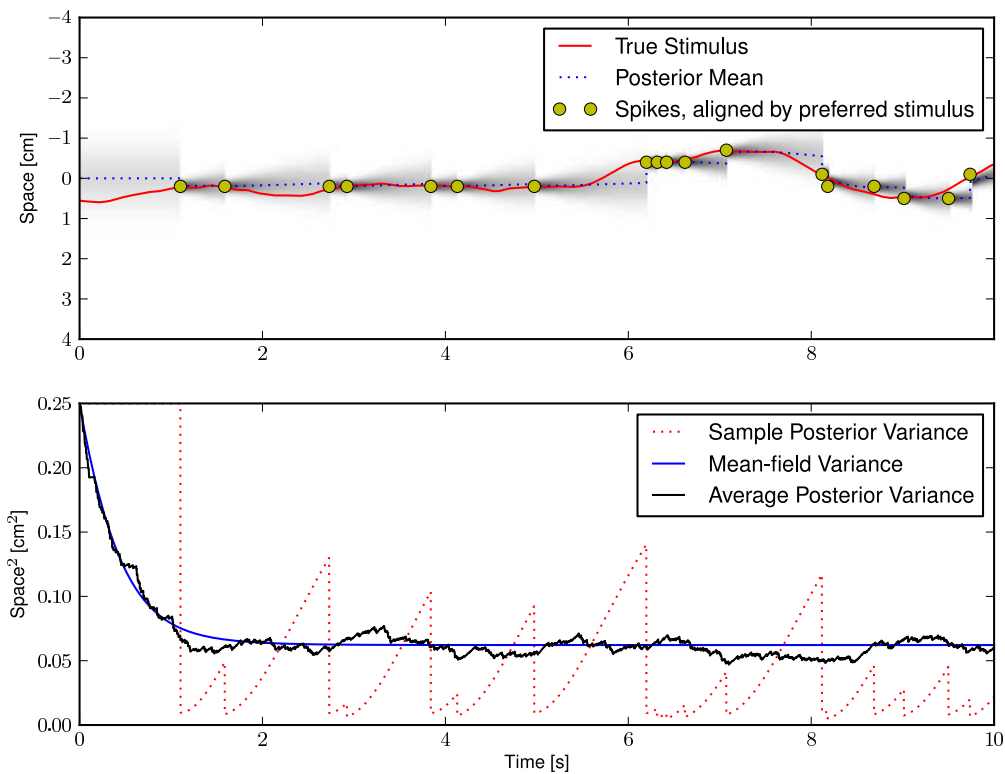}
\caption{Inference task for the $P=2$ process. The upper panel shows the true stimulus or signal $x(t)$ in red solid line, the filtered posterior mean in the blue dotted line and the shading represents the posterior probability distribution. The lower panel shows the evolution of the posterior variance. The red dotted line gives the sample variance of the sample in the upper panel, the solid black line shows an average over 100 realizations and the blue line gives the evolution of the mean-field equation with the same initial conditions.}
\label{fig:3}
\end{figure}

In section \ref{sec:ou} we considered Markovian Gaussian processes. However, within an ecologically oriented setting, it is often more natural to assume smooth priors \cite{Huys2007}. Since in general smooth priors lead to non-Markovian processes, which are very hard to deal with mathematically, we look at a special case of a smooth non-Markovian prior. Specifically, we will look at the multidimensional embedding $\mathbf{X}(t)$ of the smooth processes given by \eref{eqn:vector_sde}.
We choose to study these processes as they allow us to consider smooth stochastic processes with the same set of parameters
as were used to consider the OU process. \Fref{fig:3} shows the general inference framework for higher-order smooth processes.
%Another interesting aspect is that we can take a limit to
%infinite-order stochastic processes (stochastic processes with infinite continuous derivatives, usually
%generated by the radial-basis-function kernel \cite{Rasmussen2006}) by considering the limit
%$$
%\lim_{\alpha\to\infty} \left(1+ \frac{x^2}{2\alpha}\right)^{-\alpha} = \rme^{-x^2/2}.
%$$
%Defining $\gamma = \sqrt{P } g$, $\eta^2 = h^2 (g \sqrt{P})^{P}$ we will have
%\begin{equation}
%\lim_{P\to\infty} \left<\tilde{x}(\omega) \tilde{x}^*(\omega)\right> = h^2 \exp\left(-4\pi^2 \omega^2/g\right),
%\end{equation}
%which gives us the known RBF kernel.
\par

Taking the observation matrix $A_{ij} = \delta_{1,i}\delta_{1,j} \alpha^2$ with
pseudo-inverse $A^+ = \delta_{1,i}\delta_{1,j} /\alpha^2$, the form of $A$ will allow us to somewhat simplify equation \eref{eqn:var_matrix_jump} and we can write a simple mean-field equation for the evolution of $\left<\Sigma\right>_t = \bar{\Sigma}(t)$. We have in the mean-field approximation
\begin{equation}
\frac{\rmd \bar\Sigma(t)}{\rmd t} = -\Gamma \bar\Sigma(t) - \bar\Sigma(t) \Gamma^\top + H^2 -\lambda \ \frac{\bar\Sigma(t) A^+ \bar\Sigma(t) }{1+\bar\Sigma(t)_{1,1}/\alpha^2}~.
\label{eqn:dynamics_mf_matrix}
\end{equation}
This equation still describes the average performance of the Bayesian filter remarkably well, not only in the equilibrium phase but also in the transient period (see \fref{fig:3}). Note that the sole contributor to the nonlinearity in the equation above is the first column of the
covariance matrix. Given $\bar\Sigma_{1,j}$ solving for the remaining elements of the matrix is a matter of
linear algebra.
\par
\begin{figure}
\includegraphics[width=\columnwidth]{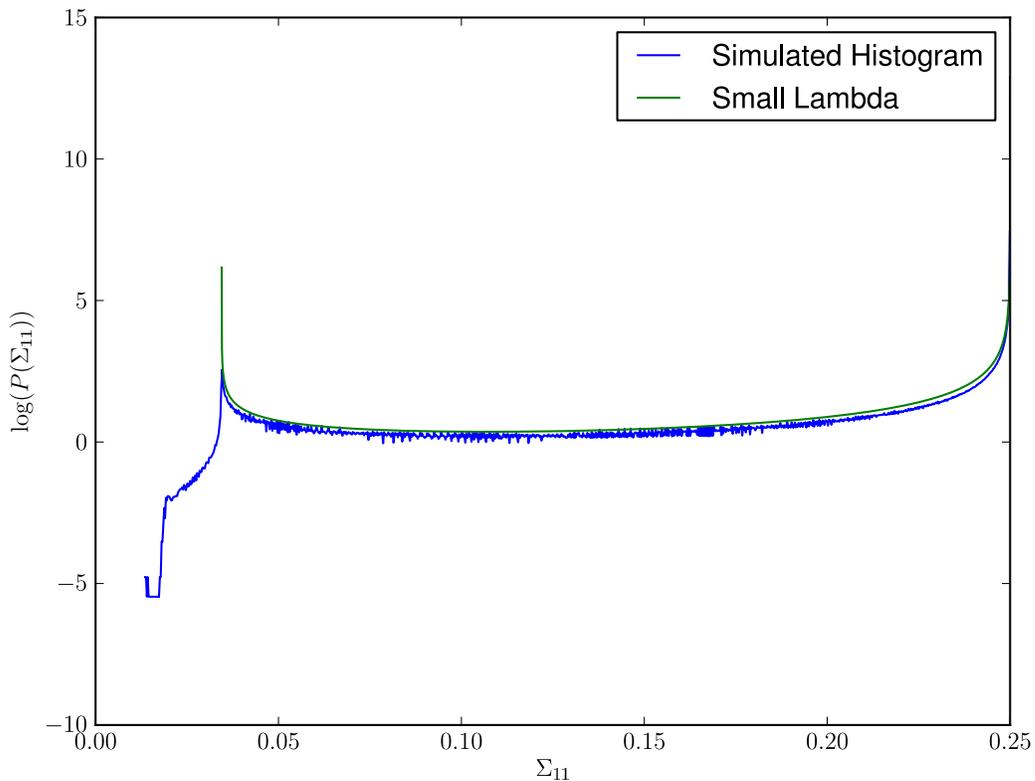}
\caption{The small firing rate approximation provides a very good account of the distribution of the variance of the observed process $\Sigma_{11}$.}
\label{fig:4}
\end{figure}

The argument made in \ref{sec:app_bound} cannot be rigorously be made here, as we have $\Sigma$ evolving
in a high dimensional space with nonlinear jumps. In principle, one could map out a subspace such
that $P(\Sigma,t)$ is zero outside of it, and then proceed to locate a set of $\Sigma$ such that
$(\Sigma^{-1}-A^+)^{-1}$ falls in it, and then seek to solve equation \eref{eqn:DCKE_multidimensional} in that
domain. This could, in principle, be made rigorous, but as we will see below, this approach is not as useful
when considering smoother processes. To see why, let us study the small $\lambda$ limit as we did in
the OU process. Assuming $\lambda \ll \nabla \cdot B$, we can assume that the covariance matrix has
nearly relaxed to its equilibrium value $\Sigma_0$ every time there is a spike. Whenever there is a spike
the covariance matrix jumps to $\Sigma' = (\Sigma_0^{-1} + A^+)^{-1}$. The free evolution is then given
by
$$
\Sigma(t) = \rme^{-(t-t_0)\Gamma}\Sigma'\rme^{-(t-t_0)\Gamma^\top} + \int_{t_0}^t \rmd s\, \rme^{-(t-s)\Gamma} H^2 \rme^{-(t-s)\Gamma^\top}.
$$
Changing variables to $\tau = t-t_0$ we have
$$
\Sigma(\tau) = \rme^{-\tau\Gamma}\Sigma'\rme^{-\tau\Gamma^\top} + \int_{0}^{\tau} \rmd s\, \rme^{-s\Gamma} H^2 \rme^{-s\Gamma^\top}.
$$
Clearly there is no simple solution to these matrix equations as there is for the OU case, but we can still
use the same approach as before numerically. We can assume an unknown function of $\Sigma$ that
gives us the time since the spike $\tau(\Sigma)$. 
%Though we cannot evaluate it analytically, this is simply obtainable through simulations. 
We can then proceed as before and evaluate the marginal distributions.
We would have for $\Sigma_{1,1}$
$$
P(\Sigma_{11}) =\frac{ P(\tau(\Sigma_{11}))}{ \left|\frac{\rmd\Sigma_{11}}{\rmd\tau}\right|}\propto \frac{\rme^{-\lambda \tau}}{\left|\frac{\rmd\Sigma_{11}}{\rmd\tau}\right|}.
$$
When we analyzed the analog of this relation for the OU process we could find a simple correspondence
between the divergence of the distribution around the equilibrium value of the variance and the
parameters $\lambda$ and $\gamma$. Though we can argue similarly in this case, note that
immediately after the jump, we have $\rmd\Sigma_{11}/\rmd\tau = \Sigma'_{12} = 0$ which results itself in a divergence
in the distribution in this approximation. So, even in the limit $\lambda \ll \nabla \cdot B$, there is always
some probability mass concentrated around the value $\Sigma'$, so the divergence argument is not
so strong for smoother processes. This can be seen in \fref{fig:4}.
\par

\subsection{Prediction Error}
%Note that it is straightforward to derive 
The prediction error can easily be derived as a function of the filtering error in this framework. We have the predictive probability
$P(\mathbf{X}_{t+\delta}|\{N^m([0,t])\})$, with $\delta>0$. The average prediction error would simply
be the deviation of $\mathbf{X}_{t+\delta}$ from the estimator $\mu(t+\delta)$.
The prediction error is then
\begin{equation}
\mathcal{P}(\delta) = \left<(\mathbf{X}(t+\delta)-\mu(t+\delta))(\mathbf{X}(t+\delta)-\mu(t+\delta))^\top\right>_{\mathbf{X},\{N^m([0,t])\}}.
\label{eqn:C_delta}
\end{equation}
This gives us the matrix $\mathcal{S}(t)$ when $\delta= 0$. For $\delta>0$ we have the prediction error matrix.
Note that given a value of $\mathbf{X}(t)$ and a realization of the Wiener process $\rmd \mathbf{W}(s)$ for $t\leq s \leq t+\delta$, we have $\mathbf{X}(t+\delta) = \int_t^{t+\delta} \rme^{-(t+\delta-s)\Gamma} H \rmd \mathbf{W}(s) + \rme^{-\delta\Gamma}\mathbf{X}(t)$. Clearly, conditioning on $\mathbf{X}(t)$ the above average is only over the Wiener process between $t$ and $t+\delta$. The estimator $\mu(t+\delta)$ is also given by $\mu(t+\delta) = \rme^{-\delta \Gamma} \mu(t)$ in the absence of spikes. The prediction error matrix will then be given by
\begin{eqnarray}
\mathcal{P}(\delta)  =& \left<( \int_t^{t+\delta} \rme^{-(t+\delta-s)\Gamma} H \rmd \mathbf{W}(s) + \rme^{-\delta\Gamma}\mathbf{X}(t)-\rme^{-\delta\Gamma}\mu(t))\times\right.\nonumber
\\
&\left. (\int_t^{t+\delta} \rme^{-(t+\delta-u)\Gamma} H \rmd \mathbf{W}(u) + \rme^{-\delta\Gamma}\mathbf{X}(t)-\rme^{-\delta\Gamma}\mu(t))^\top \right>_{\mathbf{X},\{N^m([0,t])\}}.
\end{eqnarray}
Since $ \rme^{-(t+\delta-u)\Gamma} H$ is non-anticipating and does not depend on $\mathbf{X}(t)$ or $N^m(t)$, we have that (see \cite{Gardiner2004})
$$\left<\int_{t}^{t+\delta} \rme^{-(t+\delta-u)\Gamma} H\rmd \mathbf{W}(u) \int_t^{t+\delta} (\rme^{-(t+\delta-s)\Gamma} H\rmd \mathbf{W}(s))^\top \right> = \int_t^{t+\delta} \rme^{-(t+\delta-u)\Gamma}H^2 \rme^{-(t+\delta-u)\Gamma^\top} \rmd u$$
and therefore, changing variables,
\begin{equation}
\mathcal{P}(\delta)= \rme^{-\delta\Gamma} \mathcal{P}(0) \rme^{-\delta \Gamma^\top} + \int_{0}^\delta \rme^{-s \Gamma} H^2 \rme^{-s\Gamma^\top}\rmd s.
\label{eqn:pred_error}
\end{equation}
This is the usual relation for the evolution of the variance of linear stochastic processes, and it shows us that the prediction error is a simple function of the filtering error. This is also a consequence of the Markov nature of the posterior probability. Taking a non-Markov prior process would result in a posterior probability whose parameters could not be described by a set of ordinary differential equations.\par
\begin{figure}
\includegraphics[width=.9\columnwidth]{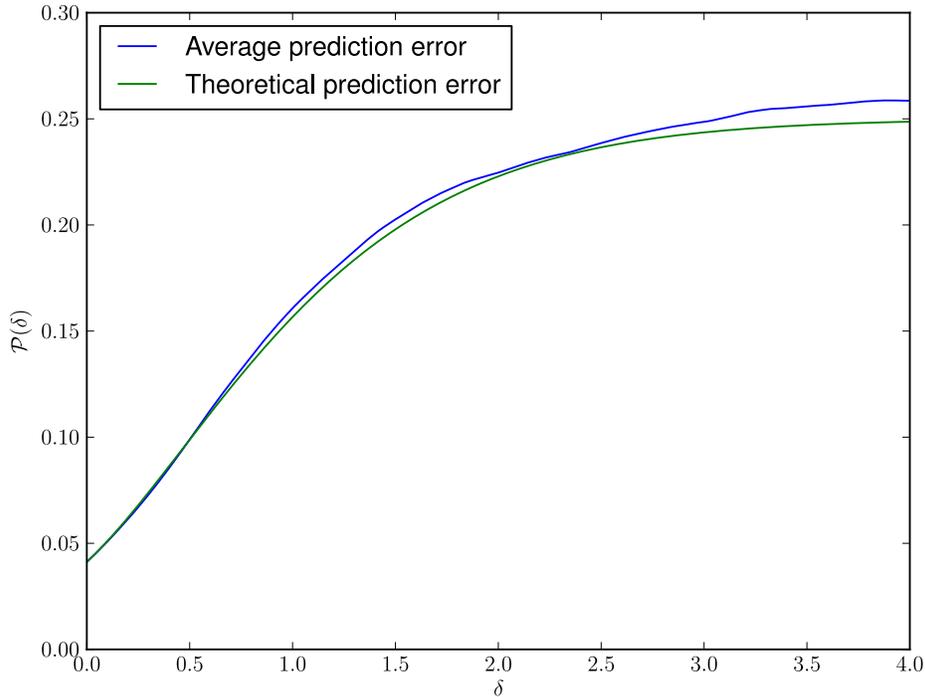}
\caption{The evolution of the average prediction error $\mathcal{P}(\delta)$ is completely determined by the filtering error $\mathcal{P}(0)$. The blue line shows the prediction error obtained from the optimal filter in simulations, whereas the green line shows the evolution of the prediction error according to \eref{eqn:pred_error} with the initial condition given by the average filtering error obtained in the simulations. The small discrepance between both curves is due to finite sample size effects.}
\label{fig:5}
\end{figure}

In \fref{fig:5} we show the comparison between the theoretical result in \eref{eqn:pred_error} and simulation results for the prediction error. We can see that the prediction error is very well described by the derived equation.

\section{Optimal Population Coding}
\label{optimal}
We can now apply the results from the previous section to study the optimal coding strategies for
neurons in the dynamic case. The study of bell-shaped and more specifically Gaussian tuning curves
has been frequently discussed in the literature \cite{Zhang1999a,Bethge2002,Yaeli2010,Berens2011}.
Our framework treats the case of densely packed Gaussian tuning functions for stochastic dynamic
stimuli. We can study the performance of the tuning functions by mapping the minimal mean squared
error for an encoder given by a certain tuning function. In our case, given the sensory neurons, specified by their
preferred stimuli $\Theta_m$, the tuning functions are specified by the maximal firing rate $\phi$ and the
tuning width $\alpha$.\par

\begin{figure}
\includegraphics[width=\columnwidth]{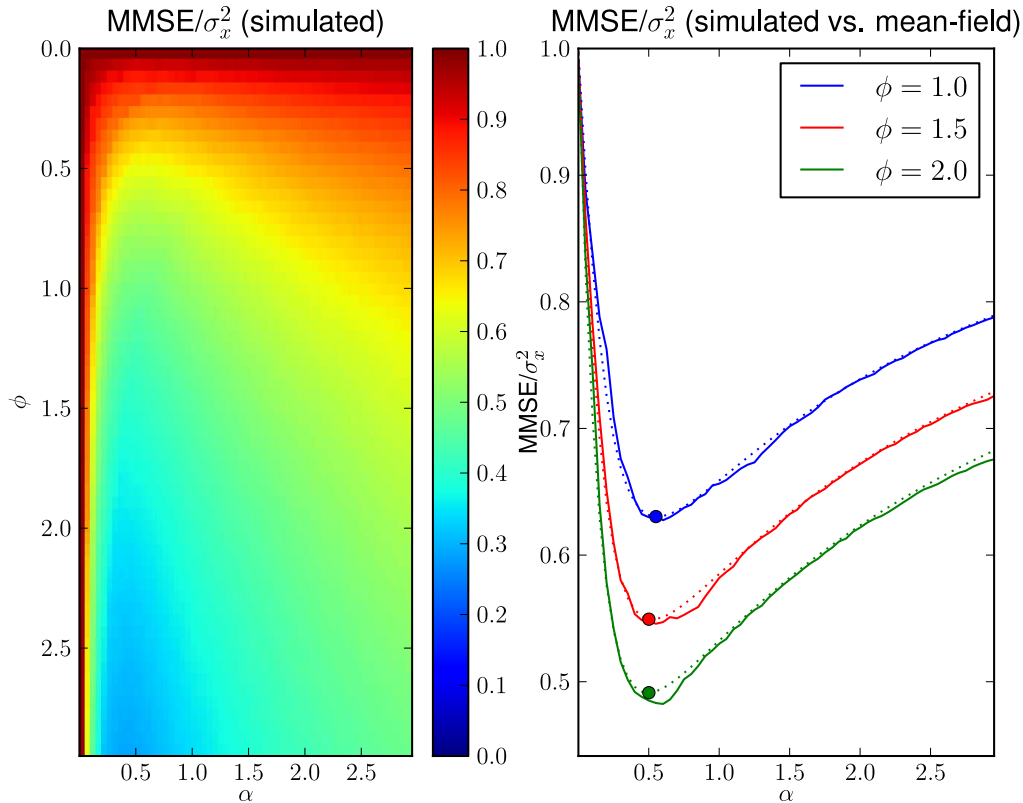}
\caption{MMSE landscape as a function of the encoder's tuning width $\alpha$ and maximal firing rate $\phi$ for the one-dimensional OU process. For simplicity we show the case where $\gamma = \eta =1$. When either parameters are very near $0$, the MMSE is equal to the stimulus' equilibrium variance. For fixed values of $\phi$ we observe the existence of a finite optimal tuning width, as is shown in the right panel.}
\label{fig:6}
\end{figure}

\fref{fig:6} shows a colormap of the MMSE for an
Ornstein-Uhlenbeck process for a range of values of $\alpha$ and $\phi$. There are some clear trends
to be observed. First, we note that for any given value of $\alpha$, increasing the maximal firing rate
$\phi$ leads, not surprisingly, to a decrease in the MMSE. When the value of $\phi$ tends to zero, the MMSE tends to
the equilibrium variance of the observed process, which is expected, as in the absence of spikes, the
best estimate of the stimulus is given by the equilibrium distribution of the OU process. A second
interesting observation is that for any given fixed value of $\phi$ there is a finite value of $\alpha$ which
minimizes the MMSE. This is in accord with findings in \cite{Yaeli2010,Berens2011,Bethge2002},
which also report that the optimal tuning width is a finite value, as well as with experimental data \cite{Benucci2009,Gutfreund2006}. This can be intuitively understood by
noting that when we decrease the tuning width keeping $\phi$ constant, we decrease the population's
firing rate $\lambda$, and eventually, for very sharp tuning functions, we will have a vanishingly small number of spikes.
On the other hand, increasing the width $\alpha$ will lead to a gain in information through more spikes
but eventually these spikes become so uninformative that there is no more advantage from increasing
$\alpha$, as these barely give us more information than is already present in the equilibrium distribution. The right panel of \fref{fig:6} shows the MMSE as a function of $\alpha$
for some values of $\phi$.\par

\begin{figure}
\includegraphics[width=\columnwidth]{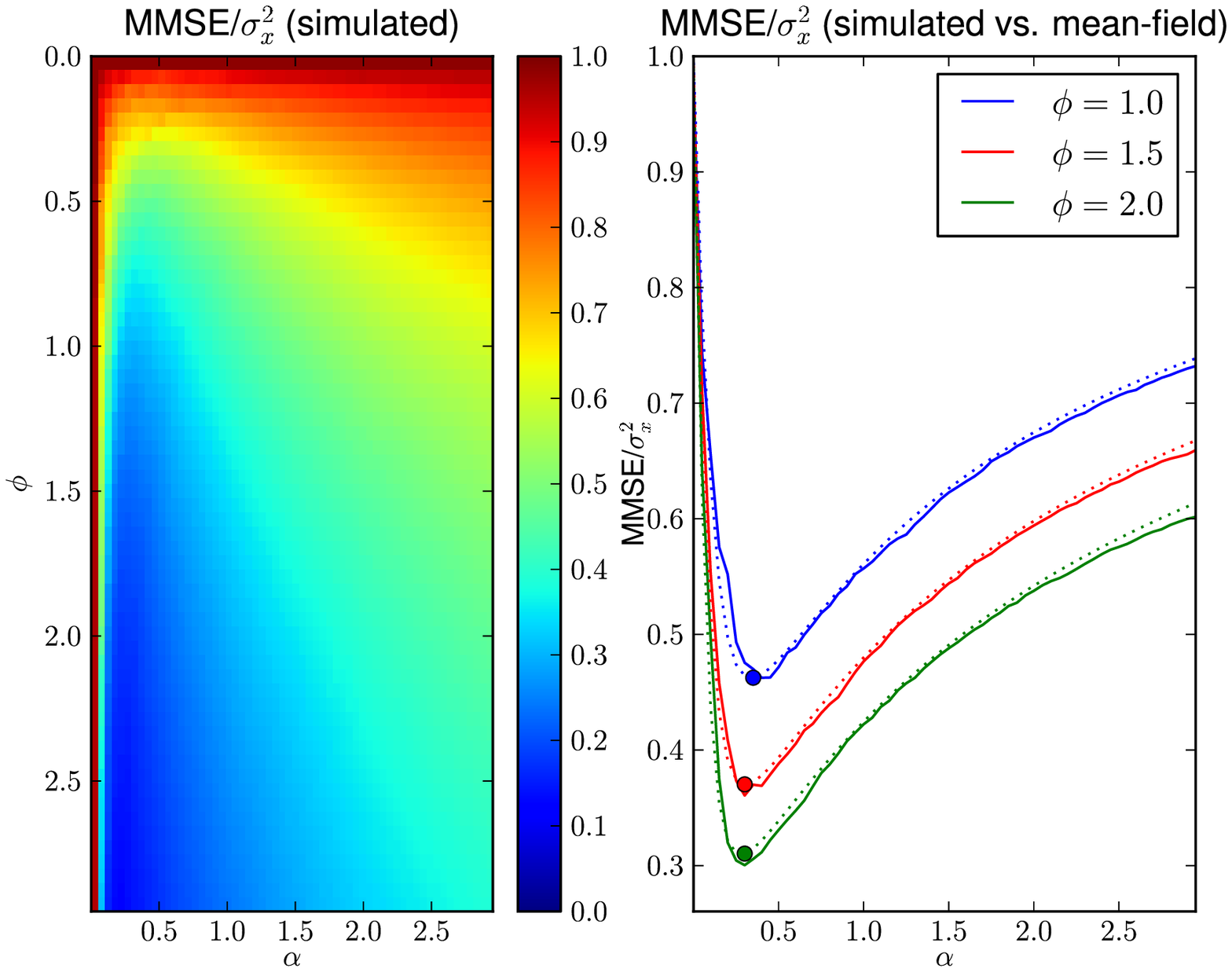}
\caption{MMSE landscape as a function of the encoder's tuning width $\alpha$ and maximal firing rate $\phi$ for the second-order OU process. Again, we show the case where $\gamma = \eta =1$. When either parameters are very near $0$, the MMSE is equal to the stimulus' equilibrium variance. For fixed values of $\phi$ we observe the existence of a finite optimal tuning width, as is shown in the right panel.}
\label{fig:7}
\end{figure}

The situation is very similar when we consider higher-order processes. Although the decay of the
posterior covariance to its equilibrium value follows a more convoluted dynamics, the argument
relating $\alpha$ and the improvement in the reconstruction still holds. \Fref{fig:7} shows the analog of \fref{fig:6} for the process with $P=2$. We can see that
the general structure of the color plot is very similar. One important difference
though, is that for the same firing rate, we are able to obtain a larger improvement in the MMSE, relative
to the equilibrium variance of the process. This is clear in the right panel of \fref{fig:7}.
This is to be expected, as the higher-order processes have more continuity constraints and are therefore
more \emph{predictable}.

\subsection{Ecological Analysis of the Optimal Code}

We will now look at how the optimal value of $\alpha$ depends on the parameters of the process. We can, fixing $\phi,\, \gamma$ and $\eta$ find the optimal value of $\alpha$ that minimizes
the MMSE. This will give us a mapping of how the optimal tuning width depends on the different
parameters of the stimulus. We consider the case of the second-order Ornstein-Uhlenbeck process ($P=2$). Note that changing $\gamma$ also influences the signal variance, as the equilibrium variance of the stimulus is given by $\sigma^2_x = \eta^2/4\gamma^3$. To better separate these influences of the timescale and variance of the stimulus, we chose to look at the process given by an OU process with diffusion coefficient $\eta' = \eta\gamma^{3/2}$ resulting in a equilibrium variance of $\sigma^2_x = \eta^2/4$. In that way we can better analyze the influence of the timescale and intensity of the covariance of the stimulus.\par

\begin{figure}
\includegraphics[width=\columnwidth]{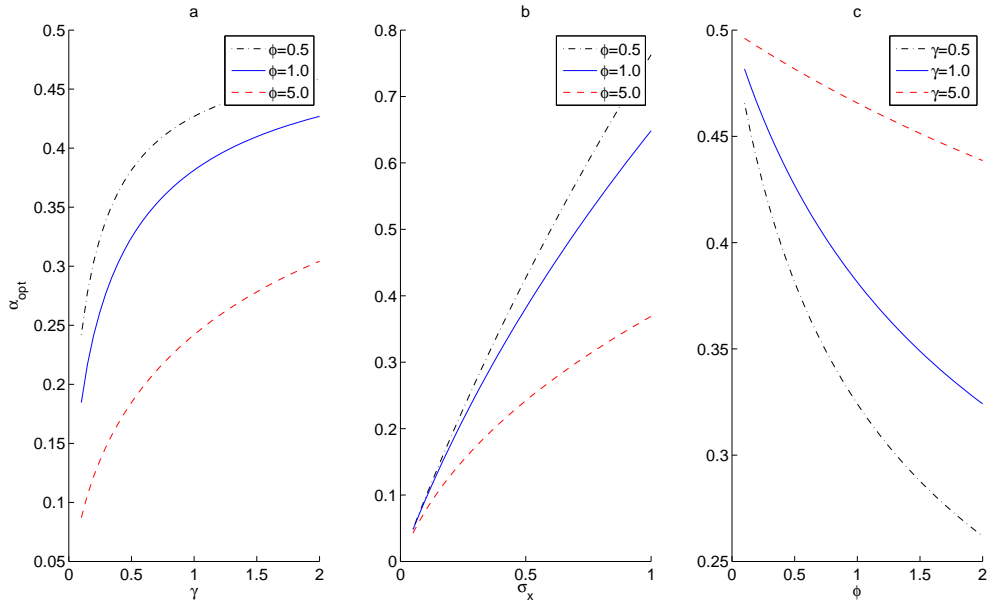}
\caption{Dependence of the optimal tuning width on the parameters of the stimulus and encoder for the $P=2$ process. Panel a shows the optimal tuning width as a function of the forcing parameter $\gamma$ which is the inverse of the characteristic timescale of the stimulus for three different values of the maximal firing rate $\phi$. Panel b shows the optimal tuning width as a function of the equilibrium standard deviation of the stimulus for the same values of $\phi$. Panel c shows the optimal tuning width as a function of the maximal firing rate of the encoding neurons for three values of the inverse time constant $\gamma$.}
\label{fig:8}
\end{figure}

This analysis is shown in \fref{fig:8} for the second-order OU process. The timescale of the process strongly influences the optimal tuning width as can be seen in the figure. With an increase in $\gamma$ the optimal tuning width increases. This is in accord with the finding in \cite{Berens2011}, where it was reported that for static stimuli a larger decoding window leads to a narrower optimal tuning width. This can be cast into our framework by making an analogy between the decoding time window and the characteristic time of the decay of the autocorrelation of the observed process. A very large correlation time means a very slowly changing process, which needs less spikes and therefore allows for narrower tuning widths. This can be also understood by referring to the solution for the error distribution of the OU process given by \eref{eqn:eq_dist_sigma}. Smaller values of $\gamma$ allow for a good reconstruction of the stimulus with less spikes. An increase in the stimulus variance $\sigma^2_x$ leads to an increase in the optimal tuning width. Meanwhile, an increase in the maximal firing width of each neuron results in a decrease of the optimal tuning width, as is expected. Increasing the height of each tuning function allows the tuning functions to sharpen without decreasing the overall firing rate of the population.\par
\par

\begin{figure}
\includegraphics[width=\columnwidth]{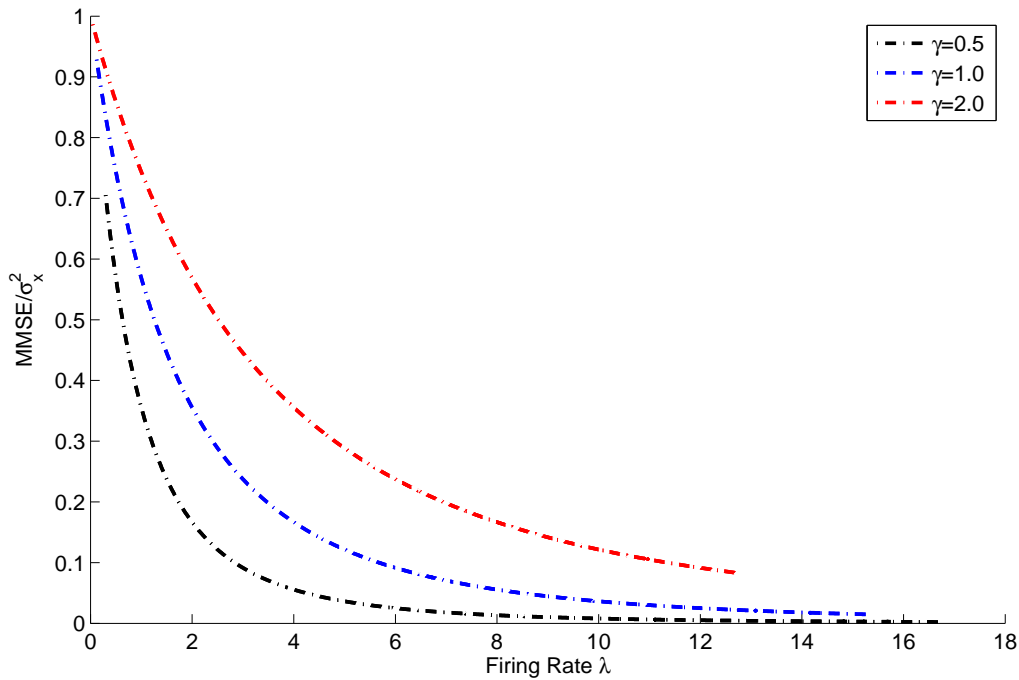}
\caption{Rate distortion curve of the optimal encoding scheme for the second-order OU process.}
\label{fig:9}
\end{figure}

An interesting perspective on the coding-decoding problem we are studying can be obtained from rate-distortion theory \cite{Cover1991}, where one seeks the optimal tradeoff between coding rate and reconstruction error (distortion). The celebrated rate-distortion curve provides  the smallest rate for which a given level of distortion can be achieved (usually asymptotically). We study the analog
 of a rate-distortion curve in our case. The rate is given by the average population firing rate $\lambda$ while the distortion would be the minimal mean squared error. In \fref{fig:9} we show a plot of the MMSE of the optimal encoder against the firing rate of the optimal code for given values of $\phi, \gamma$ and $\eta$. Interestingly, the rate-distortion curve is independent of the value of $\eta$. The dependence in $\gamma$ is as expected. For smaller values of $\gamma$ (i.e., for longer correlation times), the error decays faster with the population firing rate. Meanwhile, larger values of $\gamma$ (i.e., shorter correlation times), lead to a slower decay in the distortion as a
function of the firing rate.\par

{\Blue
\subsection{Comparison to Previous Research}

The existence of an optimal finite tuning width for unimodal tuning curves has been repeatedly determined in the literature. This has been found to be the case with the mutual information as objective function \cite{Brunel1998} as well as with the reconstruction error as an objective function \cite{Bethge2002,Berens2011,Yaeli2010,Susemihl2011a}. Our findings on the optimal tuning width as a function of the correlation time of the observed process, summarized in \fref{fig:8}, generalize the findings in \cite{Berens2011} and in \cite{Yaeli2010} for dynamic processes. Namely, it was established that longer integration times lead to narrower optimal tuning widths. In our case, longer integration times correspond to longer correlation times and therefore smaller values of $\gamma$. We have found that for smaller values of $\gamma$, i.e. for longer correlation times, the optimal tuning width decreases, as would be expected. Also, as was found in \cite{Yaeli2010}, we have established that noisier prior processes lead to broader optimal tuning widths (see \fref{fig:8}). These results seem to hold in a fairly general set of conditions and this indicates that the tradeoff between firing rate and accuracy is central to the coding strategies in the nervous system.
}

\section{Discussion}

The temporal aspect of neural coding has often been neglected in studies of optimal population coding. A
framework for the use of filtering in neural coding had been proposed in \cite{Huys2007}, but even there
the offline paradigm was favored, and most studies either bypass the issue of time completely
\cite{Tkacik2010} or resort to time-binning to address the temporal aspects \cite{Berens2011,Gerwinn2009}.
We have extended the filtering theory presented in \cite{Bobrowski2009,Yaeli2010} for finite-state
Markov processes and static processes, where the whole spike train is used for decoding and, using a model
of the dynamics of the stimulus, we have addressed the decay of the information as well.\par
Our framework generalizes a number of findings from previous studies. The finding from \cite{Berens2011}
and \cite{Yaeli2010} that the reconstruction error decays with the length of the decoding time-window is
here framed in terms of the correlation time of the stimulus. We also find the same relation between the
optimal tuning width for bell-shaped tuning functions and the correlation time as had been found for the
tuning width as a function of the decoding time.\par

Another advantage of our treatment is that it allows us to study more complex aspects of temporal correlation
than just the length of the decoding window. By considering a class of higher-order stochastic processes, we
can address different temporal correlation structures. This had been also proposed in \cite{Huys2007}, but
not explored any deeper. In that spirit,
we have established that smoother processes allow for a more efficient coding strategy with the same firing
rate in the \emph{sensory} layer. The complexity of the process results in a higher cost in the decoding
of the spike train, however, implying that although smooth processes allow for very efficient reconstruction
strategies, they will require a more complex decoding strategy. In
\cite{Natarajan2008}, a decoding strategy using recoding of spikes for RBF processes had been
suggested. The approach taken in \cite{Bobrowski2009} is to encode the dynamics of the process explicitly
into the recurrent connections of the decoding network, which would be straightforward to generalize for
smooth processes, but would imply that the number of decoding neurons scales with the order of the process.
Clearly, biological systems are not concerned with the degree of continuity, and the model of the temporal structure
of the world should be adapted to the (biological and evolutionary) experience of the natural environment.
This does, however, suggest a tradeoff between the effort spent in optimally coding a stimulus and the
effort spent in decoding it.\par
Our approach allows for a more flexible and structured way of dealing with the temporal aspects in the neural
code. The generalization of the filtering scheme for more complex neuron models such as generalized
linear models is relatively simple, although the assumption of dense tuning functions clearly cannot hold
then and the solution of the filtering problem will not be Gaussian anymore. One can still consider a
Gaussian approximation to it, as has been done in \cite{Pfister2010}, or resort to particle filter approximations \cite{Oxford2011}. One other interesting direction for
further research is to develop filtering approaches to biologically inspired neuron models, such as
leaky-integrate-and-fire models. The hurdles in this case are the same, the Gaussian assumption will not
hold and we have to work with approximations. However, through a systematic treatment of the resulting
problems, more insights might be gained into the temporal aspects of the neural code.\par

\section{Acknowledgements}

The work of A.S. was supported by the DFG Research Training Group GRK 1589/1. The work of R.M. was ...

We kindly thank...

\section{References}
\bibliography{library}{}
\bibliographystyle{unsrt}

\appendix

\section{Definition of $\Gamma$ and $H$}
\label{sec:app_matrix}
To obtain the process defined in  \eref{eqn:langevin_eqn} we define
$$
\Gamma_{i,j} = -\delta_{i+1,j} + \delta_{P,i} {P\choose j-1} \gamma^{P-j+1},\textrm{ and }  H_{i,j} = \delta_{i,P}\delta_{j,P}\eta.
$$

\section{Definition of the Pseudo-determinant}
\label{sec:app_determ}
The pseudo-determinant of a square $n\times n$ matrix $A$ is given by
$$
\textrm{det}^*(A) = \lim_{\alpha\to 0} \frac{\det(A+\alpha I)}{\alpha^{n-rank(A)}}.
$$
For positive semi-definite matrices as are used in the text, the pseudo-determinant is the product of all
non-zero eigenvalues.

\section{Boundary Conditions for the Differential Chapman-Kolmogorov Equation}

\label{sec:app_bound}

We want to show that in the equilibrium $P( s)_{eq} = 0,\,\forall  s>\eta^2/2 s$. We will
proceed by cases. If $\alpha^2<\eta^2/2\gamma$, we have that the jump term will be absent of
\eref{eqn:DCKE_onedimensional} for $  s>\eta^2/2 s$ and we will have
$$
\frac{\partial P( s,t)}{\partial t} = \frac{\partial}{\partial  s}\left( (2\gamma  s - \eta^2)P( s,t)\right) -\lambda P( s,t).
$$
It is easy to see that given a solution of equation
$$
\frac{\partial P^*( s,t)}{\partial t} = \frac{\partial}{\partial  s}\left( (2\gamma  s - \eta^2)P^*( s,t)\right)
$$
with some initial condition $P^*( s,t_0) = P_0( s)$,
$P( s,t) = \rme^{-\lambda (t-t_0)} P^*( s,t)$ is a solution of the first equation with the same
initial condition. Therefore
$P( s,t) < P^*( s,t),\, t>t_0$. But $P^*( s,t)$ is the solution of the Liouville equation for the system
$\dot{ s} = -2\gamma  s + \eta^2$. Namely, as $t \to \infty$, $P^*( s,t) \to \delta( s-\eta^2/2\gamma)$. Therefore, $\forall  s > \eta^2/2\gamma$, $P( s,t)\to 0$, as $t \to \infty$.\par
If $\alpha^2>\eta^2/2\gamma$, we first proceed in the same manner, but taking $ s >\alpha^2$ and
imposing an absorbing boundary condition at $ s = \alpha^2$.
Thus we show that $P( s,t)_{eq} = 0, \, \forall  s>\alpha^2$. We can then subsequently
apply the same argument for the intervals $[j^n(\alpha^2),j^{n-1}(\alpha^2)]$, as long as
$j^n(\alpha^2)>\eta^2/2\gamma$. Finally, we use the same argument for $
[\eta^2/2\gamma,j^m(\alpha^2)]$, where $m$ is the highest integer such that $j^m(\alpha^2)>
\eta^2/2\gamma$. This shows our desired result.

\end{document}